\documentclass[showpacs,nofootinbib,tightenlines,amsmath,amssymb]{revtex4}
\usepackage{bm}
\usepackage{times}
\usepackage{graphicx}
\usepackage{dcolumn}
\begin{document}

\title{Energy conditions, traversable wormholes and dust shells}

\author{Francisco S. N. Lobo}
\email{flobo@cosmo.fis.fc.ul.pt}

\affiliation{Centro de Astronomia
e Astrof\'{\i}sica da Universidade de Lisboa, \\
Campo Grande, Ed. C8 1749-016 Lisboa, Portugal}


\begin{abstract}
Firstly, we review the pointwise and averaged energy conditions,
the quantum inequality and the notion of the ``volume integral
quantifier'', which provides a measure of the ``total amount'' of
energy condition violating matter. Secondly, we present a specific
metric of a spherically symmetric traversable wormhole in the
presence of a generic cosmological constant, verifying that the
null and the averaged null energy conditions are violated, as was
to be expected. Thirdly, a pressureless dust shell is constructed
around the interior wormhole spacetime by matching the latter
geometry to a unique vacuum exterior solution. In order to further
minimize the usage of exotic matter, we then find regions where
the surface energy density is positive, thereby satisfying all of
the energy conditions at the junction surface. An equation
governing the behavior of the radial pressure across the junction
surface is also deduced. Lastly, taking advantage of the
construction, specific dimensions of the wormhole, namely, the
throat radius and the junction interface radius, and estimates of
the total traversal time and maximum velocity of an observer
journeying through the wormhole, are also found by imposing the
traversability conditions.

\end{abstract}


\maketitle

\section{Introduction}

Much interest has been aroused in wormholes since the
Morris-Thorne article \cite{Morris}. These act as tunnels from one
region of spacetime to another, possibly through which observers
may freely traverse. Wormhole physics is a specific example of
solving the Einstein field equation in the reverse direction,
namely, one first considers an interesting and exotic spacetime
metric, then finds the matter source responsible for the
respective geometry. In this manner, it was found that these
traversable wormholes possess a peculiar property, namely exotic
matter, involving a stress-energy tensor that violates the null
energy condition \cite{Morris,MTY,MV}. In fact, they violate all
the known pointwise energy conditions and averaged energy
conditions, which are fundamental to the singularity theorems and
theorems of classical black hole thermodynamics. The weak energy
condition (WEC) assumes that the local energy density is positive
and states that $T_{\mu\nu}U^\mu U^\nu \geq 0$, for all timelike
vectors $U^\mu$, where $T_{\mu\nu}$ is the stress energy tensor.
By continuity, the WEC implies the null energy condition (NEC),
$T_{\mu\nu}k^\mu k^\nu \geq 0$, where $k^\mu$ is a null vector.
Violations of the pointwise energy conditions led to the averaging
of the energy conditions over timelike or null geodesics
\cite{Tipler}. For instance, the averaged weak energy condition
(AWEC) states that the integral of the energy density measured by
a geodesic observer is non-negative, i.e., $\int T_{\mu\nu}U^\mu
U^\nu \,d\tau \geq 0$, where $\tau$ is the observer's proper time.
Although classical forms of matter are believed to obey these
energy conditions, it is a well-known fact that they are violated
by certain quantum fields, amongst which we may refer to the
Casimir effect.

Pioneering work by Ford in the late 1970's on a new set of energy
constraints \cite{ford1}, led to constraints on negative energy
fluxes in 1991 \cite{ford2}. These eventually culminated in the
form of the Quantum Inequality (QI) applied to energy densities,
which was introduced by Ford and Roman in 1995 \cite{F&R1}. The QI
was proven directly from Quantum Field Theory, in four-dimensional
Minkowski spacetime, for free quantized, massless scalar fields
and takes the following form
\begin{equation}
\frac{\tau_0}{\pi}\int_{-\infty}^{+\infty}\frac{
\langle{T_{\mu\nu}U^{\mu}U^{\nu}}\rangle}
{\tau^2+{ \tau^2_0}}d\tau\geq-\frac{3}{32\pi^2\tau^4_0}, \label{1}
\end{equation}
in which, $U^\mu$ is the tangent to a geodesic observer's
wordline; $\tau$ is the observer's proper time and $\tau_0$ is a
sampling time. The expectation value $\langle\rangle$ is taken
with respect to an arbitrary state $|\Psi\rangle$. Contrary to the
averaged energy conditions, one does not average over the entire
wordline of the observer, but weights the integral with a sampling
function of characteristic width, $\tau_0$. The inequality limits
the magnitude of the negative energy violations and the time for
which they are allowed to exist.

The basic applications to curved spacetimes is that these appear
flat if restricted to a sufficiently small region. The application
of the QI to wormhole geometries is of particular interest
\cite{F&R2}. A small spacetime volume around the throat of the
wormhole was considered, so that all the dimensions of this volume
are much smaller than the minimum proper radius of curvature in
the region. Thus, the spacetime can be considered approximately
flat in this region, so that the QI constraint may be applied. The
results of the analysis is that either the wormhole possesses a
throat size which is only slightly larger than the Planck length,
or there are large discrepancies in the length scales which
characterize the geometry of the wormhole. The analysis imply that
generically the exotic matter is confined to an extremely thin
band, and/or that large red-shifts are involved, which present
severe difficulties for traversability, such as large tidal forces
\cite{F&R2}. Due to these results, Ford and Roman concluded that
the existence of macroscopic traversable wormholes is very
improbable (see \cite{Roman} for an interesting review). It was
also shown that, by using the QI, enormous amounts of exotic
matter are needed to support the Alcubierre warp drive and the
superluminal Krasnikov tube \cite{P&F2,E&R,LoboSLT}. However,
there are a series of objections that can be applied to the QI.
Firstly, the QI is only of interest if one is relying on quantum
field theory to provide the exotic matter to support the wormhole
throat. But there are classical systems (non-minimally coupled
scalar fields) that violate the null and the weak energy
conditions \cite{B&V}, whilst presenting plausible results when
applying the QI. Secondly, even if one relies on quantum field
theory to provide exotic matter, the QI does not rule out the
existence of wormholes, although they do place serious constraints
on the geometry. Thirdly, it may be possible to reformulate the QI
in a more transparent covariant notation, and to prove it for
arbitrary background geometries.

More recently, Visser {\it et al} \cite{VKD,Kar2}, noting the fact
that the energy conditions do not actually quantify the ``total
amount'' of energy condition violating matter, developed a
suitable measure for quantifying this notion by introducing a
``volume integral quantifier''. This notion amounts to calculating
the definite integrals $\int T_{\mu\nu}U^\mu U^\nu \,dV$ and $\int
T_{\mu\nu}k^\mu k^\nu \,dV$, and the amount of violation is
defined as the extent to which these integrals become negative.
Although the null energy and averaged null energy conditions are
always violated for wormhole spacetimes, Visser {\it et al}
considered specific examples of spacetime geometries containing
wormholes that are supported by arbitrarily small quantities of
averaged null energy condition violating matter. It is also
interesting to note that by using the ``volume integral
quantifier'', extremely stringent conditions were found on ``warp
drive'' spacetimes, considering non-relativistic velocities of the
bubble velocity~\cite{LVwarp}.

As the violation of the energy conditions is a problematic issue,
depending on one's point of view \cite{B&V}, it is interesting to
note that an elegant class of wormhole solutions minimizing the
usage of exotic matter was constructed by Visser
\cite{VisserPRD,VisserNP} using the cut-and-paste technique, in
which the exotic matter is concentrated at the wormhole throat.
Using these thin-shell wormholes, a dynamic stability was
analyzed, either by choosing specific surface equations of state
\cite{Kim1,VisserPLB,Kim2}, or by considering a linearized
stability analysis around a static solution
\cite{Poisson,Eiroa,Lobolinear}. One may also construct wormhole
solutions by matching an interior wormhole to an exterior vacuum
solution, at a junction surface. In particular, a thin shell
around a traversable wormhole, with a zero surface energy density
was analyzed in \cite{LLQ}, and with generic surface stresses in
\cite{Lobo}. A similar analysis for the plane symmetric case, with
a negative cosmological constant, is done in \cite{LL}. A general
class of wormhole geometries with a cosmological constant and
junction conditions was analyzed by DeBenedictis and
Das~\cite{DeDas1}, and further explored in higher
dimensions~\cite{DeDas2}.

A particularly simple, yet interestingly enough, case is that of a
construction of a dust shell around a traversable wormhole. The
null energy condition violation at the throat is necessary to
maintain the wormhole open, although the averaged null energy
condition violating matter can be made arbitrarily small
\cite{VKD,Kar2}. Thus, in order to further minimize the usage of
exotic matter, one may impose that the surface stress energy
tensor obeys the energy conditions at the junction surface. For a
pressureless dust shell, one need only find the regions in which
the surface energy density is non-negative, to determine the
regions in which all of the energy conditions are satisfied.

The plan of this paper is as follows: In section II, we present a
specific spacetime metric of a spherically symmetric traversable
wormhole, in the presence of a generic cosmological constant,
analyzing the respective mathematics of embedding. We verify that
the null energy and the averaged null energy conditions are
violated, as was to be expected. Using the ``volume integral
quantifier'' and considering the specific example of the Ellis
``drainhole'', we verify that the construction of this spacetime
geometry can be made with arbitrarily small quantities of averaged
null energy condition violating matter. In section III, we present
the unique exterior vacuum solution. In section IV, we construct a
pressureless dust shell around the interior wormhole spacetime, by
matching the latter to the exterior vacuum solution. We also
deduce an expression governing the behavior of the radial pressure
across the junction surface. In section V, we find regions where
the surface energy density is positive, thereby satisfying all of
the energy conditions at the junction, in order to further
minimize the usage of exotic matter. In section VI, specific
dimensions of the wormhole, namely, the throat radius and the
junction interface radius, and estimates of the total traversal
time and maximum velocity of an observer journeying through the
wormhole, are also found by imposing the traversability
conditions. Finally, we conclude in section VII.

\section{Interior wormhole solution}

Consider the following static and spherically symmetric line
element (with $G=c=1$)
\begin{equation}
ds^2=-e ^{2\Phi(r)}\,dt^2+\left(\frac{\Lambda}{3}r^2-
\frac{m(r)}{r}\right)^{-1}\,dr^2+r^2 \,(d\theta ^2+\sin ^2{\theta}
\, d\phi ^2) \label{metricwormhole},
\end{equation}
where $\Phi(r)$ and $m(r)$ are arbitrary functions of the radial
coordinate, $r$, and $\Lambda$ is the cosmological constant.
$\Phi(r)$ is called the redshift function, for it is related to
the gravitational redshift. We shall see ahead that this metric
corresponds to a wormhole spacetime, so that $m(r)$ can be denoted
as the form function, as it determines the shape of the wormhole
\cite{Morris}. The radial coordinate has a range that increases
from a minimum value at $r_0$, corresponding to the wormhole
throat, to $a$, where the interior spacetime will be joined to an
exterior vacuum solution.

Consider, without a significant loss of generality, an equatorial
slice, $\theta=\pi/2$, of the line element (\ref{metricwormhole}),
at a fixed moment of time and a fixed $\phi$. The metric is thus
reduced to $ds^2=(\Lambda r^2/3-m/r)^{-1}dr^2$, which can then be
embedded in a two-dimensional Euclidean space, $ds^2=dz^2+dr^2$.
The lift function, $z$, is only a function of $r$, i.e., $z=z(r)$.
Thus, identifying the radial coordinate, $r$, of the embedding
space with the slice considered of the wormhole geometry, we have
the condition for the embedding surface, given by
\begin{equation} \label{embeddingsurface}
\frac{dz}{dr}=\pm \left(\frac{1- \frac{\Lambda}{3}
r^2+\frac{m}{r}}{\frac{\Lambda}{3} r^2-\frac{m}{r}} \right)^{1/2}
\,.
\end{equation}
To be a solution of a wormhole, the radial coordinate has a
minimum value, $r=r_0$, denoted as the throat, which defined in
terms of the shape function is given by
\begin{equation}
m(r_0)=\frac{\Lambda}{3}\,r_0^3    \,. \label{defthroat}
\end{equation}
At this value the embedded surface is vertical, i.e., $dz/dr
\rightarrow \infty$. As in a general wormhole solution, the radial
coordinate $r$ is ill-behaved near the throat, but the proper
radial distance, $l(r)=\pm \int_{r_0}^{r} (\Lambda r'^2/3 -m(r')/
r')^{-1/2}\,dr'$, is required to be finite throughout spacetime.
This implies that the condition $\Lambda r'^2/3 -m(r')/ r' \geq 0$
is imposed. Furthermore, one needs to impose that the throat
flares out, which mathematically entails that the inverse of the
embedding function, $r(z)$, must satisfy $d^2r/dz^2>0$ at or near
the throat. This flaring-out condition is given by
\begin{equation}
\frac{d^2r}{dz^2}=\frac{\frac{2\Lambda}{3} r^3-m'r+m}{2
\left(r-\frac{\Lambda}{3}r^3+m \right)^2}>0 \label{flareout}\,,
\end{equation}
implying that at the throat, $r=r_0$ or $m(r_0)=\Lambda r_0^3/3$,
we have the important condition
\begin{equation}
m'(r_0)<\Lambda r_0^2   \label{flarecondition}\,.
\end{equation}
We will see below that this condition plays a fundamental role in
the analysis of the violation of the energy conditions.

Using the Einstein field equation, $G_{\hat{\mu}\hat{\nu}}+\Lambda
 g_{\hat{\mu}\hat{\nu}}=8\pi T_{\hat{\mu}\hat{\nu}}$, in an
orthonormal reference frame with the following set of basis
vectors
\begin{eqnarray}\label{basisvectors}
{\bf e}_{\hat{t}}=e^{-\Phi} \;{\bf e}_{t} \;,
\qquad
{\bf e}_{\hat{r}}=(\Lambda r^2/3- m/r)^{1/2}\; {\bf e}_{r} \;,
\qquad
{\bf e}_{\hat{\theta}}=r^{-1} \;{\bf e}_{\theta}  \;,
\qquad
{\bf e}_{\hat{\phi}}=(r \sin \theta)^{-1}\; {\bf e}_{\phi}
 \;.
\end{eqnarray}
the stress energy tensor components of the metric
(\ref{metricwormhole}) are given by
\begin{eqnarray}
\rho(r)&=&\frac{1}{8\pi}\left(\frac{1+m'}{r^2}-2\Lambda \right)  \label{rho}\,, \\
p_{r}(r)&=&\frac{1}{8\pi} \left[-\frac{r+m}{r^3}+
\frac{2\Phi'}{r}\left(\frac{\Lambda}{3}r^2-\frac{m}{r}
\right)+\frac{4\Lambda}{3} \right] \label{tau}\,, \\
p_{t}(r)&=&\frac{1}{8\pi}\left[
\left(\frac{\Lambda}{3}r^2-\frac{m}{r} \right)\left(\Phi ''+
(\Phi')^2+\frac{\Phi'}{r}\right)-
\frac{(1+r\Phi')}{2r^3}\left(m'r-m-\frac{2\Lambda}{3} r^3\right)
 +\Lambda    \right]
\label{p}\,.
\end{eqnarray}
$\rho(r)$ is the energy density, $p_r (r)$ is the radial pressure,
and $p_t(r)$ is the pressure measured in the lateral directions,
orthogonal to the radial direction.

One may readily verify that the null energy condition (NEC) is
violated at the wormhole throat. The NEC states that
$T_{\mu\nu}k^{\mu}k^{\nu}\geq0$, where $k^{\mu}$ is a null vector.
In the orthonormal frame, $k^{\hat{\mu}}=(1,1,0,0)$, we have
\begin{equation}\label{NECthroat}
T_{\hat{\mu}\hat{\nu}}k^{\hat{\mu}}k^{\hat{\nu}}=\rho(r)+p_r(r)=
\frac{1}{8\pi r^3}\,\left[m'r-m-\frac{2\Lambda}{3} r^3+
2r^2\,\Phi'\left(\frac{\Lambda}{3}r^2-\frac{m}{r}\right) \right] .
\end{equation}
Due to the flaring out condition of the throat deduced from the
mathematics of embedding, i.e., Eq. (\ref{flarecondition}), we
verify that at the throat $m(r_0)=\Lambda r_0/3$, and due to the
finiteness of $\Phi(r)$, from Eq. (\ref{NECthroat}) we have
$T_{\hat{\mu}\hat{\nu}}k^{\hat{\mu}}k^{\hat{\nu}}<0$. Matter that
violates the NEC is denoted as exotic matter.

In particular, one may consider a specific class of form functions
that impose that Eq. (\ref{metricwormhole}) is asymptotically
flat. This condition is reflected in the embedding diagram as
$dz/dr \rightarrow 0$ in the limit $l \rightarrow \pm \infty$,
i.e., $\Lambda r^2/3-m/r\rightarrow 1$ as $l \rightarrow \pm
\infty$. In this case we verify that the averaged null energy
condition (ANEC), defined as $\int
T_{\hat{\mu}\hat{\nu}}k^{\hat{\mu}}k^{\hat{\nu}} d \lambda \geq
0$, is also violated. $\lambda$ is an affine parameter along a
radial null geodesic. Carrying out an identical computation as in
\cite{MV}, we have
\begin{eqnarray}\label{ANEC}
\int T_{\hat{\mu}\hat{\nu}}k^{\hat{\mu}}k^{\hat{\nu}} d \lambda=
-\frac{1}{4\pi} \int
\frac{1}{r^2}\;e^{-\Phi}\;\sqrt{\frac{\Lambda}{3}r^2-\frac{m(r)}{r}}
\; \;dr  <0 \,.
\end{eqnarray}

One may also consider the ``volume integral quantifier'', as
defined in \cite{VKD,Kar2}, which provides information about the
``total amount'' of ANEC violating matter in the spacetime. Taking
into account Eq. (\ref{NECthroat}) and performing an integration
by parts, the volume integral quantifier is given by
\begin{eqnarray}\label{vol:int}
\int (\rho + p_r)\; dV= -\int_{r_0}^\infty \left(\Lambda r^2-m'
\right) \;\left[\ln \left(\frac{e^{\Phi}}{\Lambda r^2/3-m(r)/r}
\right) \right] \;dr   \,.
\end{eqnarray}
Consider, for simplicity, the Ellis ``drainhole'' solution
\cite{ellis} (also considered in \cite{Morris,Harris}), which
corresponds to choosing a zero redshift function, $\Phi=0$, and
the following form function
\begin{equation}\label{Homerform}
m(r)=\frac{r_0^2-r^2}{r}+\frac{\Lambda}{3}r^3 \,.
\end{equation}
Suppose now that the wormhole extends from the throat, $r_0$, to a
radius situated at $a$. Evaluating the volume integral, one
deduces
\begin{eqnarray}
\int (\rho + p_r)\; dV= \frac{1}{a} \left[\left(a^2-r_0^2 \right)
\; \ln \left(1-\frac{r_0^2}{a^2} \right) +2r_0 (r_0-a) \right] .
\end{eqnarray}
Taking the limit as $a\rightarrow r_0^+$, one verifies that $\int
(\rho + p_r)\; dV \rightarrow 0$. Thus, as in the examples
presented in \cite{VKD,Kar2}, with the form function of Eq.
(\ref{Homerform}), one may construct a traversable wormhole with
arbitrarily small quantities of ANEC violating matter. The exotic
matter threading the wormhole extends from the throat at $r_0$ to
the junction boundary situated at $a$, where the interior solution
is matched to an exterior vacuum spacetime.

\section{Exterior vacuum solution}

In general, the solutions of the interior and exterior spacetimes
are given in different coordinate systems. Therefore, to
distinguish between both spacetimes, the exterior vacuum solution,
written in the coordinate system
$(\bar{t},\bar{r},\bar{\theta},\bar{\phi})$, is given by
\begin{eqnarray}
d\bar{s}^2=-\left(1-\frac{2M}{\bar{r}}-\frac{\bar{\Lambda}}{3}
\bar{r}^2 \right) \,d\bar{t}^2
+\left(1-\frac{2M}{\bar{r}}-\frac{\bar{\Lambda}}{3} \bar{r}^2
\right)^{-1}\,d\bar{r}^2+\bar{r}^2(d\bar{\theta} ^2+\sin
^2{\bar{\theta}}\, d\bar{\phi} ^2) \,. \label{metricvacuum}
\end{eqnarray}

If $\bar{\Lambda} >0$, the solution is denoted by the
Schwarzschild-de Sitter spacetime. For $\bar{\Lambda} <0$, we have
the Schwarzschild-anti de Sitter spacetime, and of course the
specific case of $\bar{\Lambda} =0$ is reduced to the
Schwarzschild solution, with a black hole event horizon at
$\bar{r}_b=2M$. Note that the metric (\ref{metricvacuum}) is
asymptotically de Sitter, if $\bar{\Lambda} >0$ as $\bar{r}
\rightarrow \infty$, or asymptotically anti-de Sitter, if
$\bar{\Lambda} <0$, as $\bar{r} \rightarrow \infty$.

For the Schwarzschild-de Sitter spacetime, $\bar{\Lambda}
>0$, if $0<9\bar{\Lambda} M^2<1$, then the factor
$f(\bar{r})=1-2M/\bar{r}-\bar{\Lambda} \bar{r}^2/3$ possesses two
positive real roots (see \cite{Lobolinear,LLQ,Lobo} for details),
$\bar{r}_b$ and $\bar{r}_c$, corresponding to the black hole and
the cosmological event horizons of the de Sitter spacetime,
respectively. In this domain we have $2M<\bar{r}_b<3M$ and
$\bar{r}_c>3M$.

Considering the Schwarzschild-anti de Sitter metric, with
$\bar{\Lambda} <0$, the factor $f(\bar{r})$ has only one real
positive root, $\bar{r}_b$, corresponding to a black hole event
horizon, with $0<\bar{r}_b<2M$ (see \cite{Lobolinear,LLQ,Lobo} for
details).

\section{Junction conditions}

We shall match Eqs. (\ref{metricwormhole}) and
(\ref{metricvacuum}) at a junction surface, $S$, situated at
$r=\bar{r}=a$. In order for these line elements to be continuous
across the junction, $ds^2|_S=d\bar{s}^2|_S$, we consider the
following transformations
\begin{eqnarray}
\bar{t}&=&\frac{t\,e^{\Phi(a)}}{\sqrt{1-\bar{\Lambda} a^2/3-2M/a}}
\;,
\\
\frac{d\bar{r}}{dr}\Big|_{r=a}&=&\frac{\sqrt{1-\bar{\Lambda} a^2/3-2M/a}}
{\sqrt{\Lambda a^2/3-m(a)/a}} \;,  \\
\bar{\theta}&=&\theta    \quad \hbox{and} \quad
\bar{\phi}\;=\;\phi  \,.
\end{eqnarray}
$\bar{\Lambda}$ and $\Lambda$ correspond to the exterior and
interior cosmological constants, respectively, which we shall
assume continuous across the junction surface, i.e.,
$\bar{\Lambda}=\Lambda$. We shall consider that the junction
surface $S$ is a timelike hypersurface defined by the parametric
equation of the form $f(x^{\mu}(\xi^i))=0$.
$\xi^i=(\tau,\theta,\phi)$ are the intrinsic coordinates on $S$,
and $\tau$ is the proper time as measured by a comoving observer
on the hypersurface. The intrinsic metric to $S$ is given by
\begin{equation}
ds^2_{S}=-d\tau^2 + a^2 \,(d\theta ^2+\sin ^2{\theta}\,d\phi^2)
\,.
\end{equation}
Note that the junction surface, $r=a$, is situated outside the
event horizon, i.e., $a>\bar{r}_b$, to avoid a black hole
solution.

Using the Darmois-Israel formalism \cite{Darmois,Israel}, the
surface stresses at the junction interface are given by
\begin{eqnarray}
\sigma&=&-\frac{1}{4\pi a} \left(\sqrt{1-\frac{2M}{a}-
\frac{\Lambda}{3}a^2}- \sqrt{\frac{\Lambda}{3}a^2- \frac{m(a)}{a}}
\, \right)        ,
    \label{surfenergy}     \\
{\cal P}&=&\frac{1}{8\pi a} \left(\frac{1-\frac{M}{a}
-\frac{2\Lambda}{3}a^2}{\sqrt{1-\frac{2M}{a}-\frac{\Lambda}{3}a^2}}-
\zeta \, \sqrt{\frac{\Lambda}{3}a^2- \frac{m(a)}{a}} \, \right)
    \label{surfpressure}    ,
\end{eqnarray}
with $\zeta=1+a\Phi'(a)$ \cite{Lobo}. $\sigma$ and ${\cal P}$ are
the surface energy density and the tangential surface pressure,
respectively. The surface mass of the thin shell is given by
$M_{\rm s}=4\pi a^2 \sigma$. The total mass of the system, $M$, is
provided by the following expression
\begin{equation}\label{totalmass}
M=\frac{a+m(a)}{2}-\frac{\Lambda}{3}a^3+M_s
\left(\sqrt{\frac{\Lambda}{3}a^2-\frac{m(a)}{a}} - \frac{M_s}{2a}
\right) \,.
\end{equation}

In particular, considering a pressureless dust shell, ${\cal
P}=0$, from Eq. (\ref{surfpressure}), we have the following
constraint
\begin{equation} \label{zetarestriction}
\zeta \, \sqrt{\frac{\Lambda}{3}a^2-
\frac{m(a)}{a}}=\frac{1-\frac{M}{a}
-\frac{2\Lambda}{3}a^2}{\sqrt{1-\frac{2M}{a}-\frac{\Lambda}{3}a^2}}
\,,
\end{equation}
which restricts the values of the redshift parameter, $\zeta$.
Eliminating the factor containing the form function in Eq.
(\ref{surfenergy}) by using Eq. (\ref{zetarestriction}), the
surface energy density is then given by the following relationship
\begin{equation}
\sigma=\frac{1}{4\pi a \,\zeta}
\,\left[\frac{(1-\zeta)+(\zeta-\frac{1}{2})\,\frac{2M}{a}
-(2-\zeta)\frac{\Lambda}{3}a^2}{\sqrt{1-\frac{2M}{a}-
\frac{\Lambda}{3}a^2}}  \right] \,,
        \label{sigma}
\end{equation}
with $\zeta \neq 0$. This expression will be analyzed in the
following section.

It is also of interest to obtain an equation governing the
behavior of the radial pressure at the junction boundary in terms
of the surface stresses at the junction boundary \cite{MV}, given
by
\begin{eqnarray}\label{pressurebalance}
\bar{p}(a)-p(a)&=&
      \frac{1}{a}\,\left(\sqrt{1-\frac{2M}{a}-\frac{\Lambda}{3}a^2}
+\sqrt{\frac{\Lambda}{3}a^2- \frac{m(a)}{a}}\;\right)\,{\cal P}
      \nonumber       \\
&& -\left(\frac{\frac{M}{a^2}- \frac{\Lambda}{3}a}
{\sqrt{1-\frac{2M}{a}-\frac{\Lambda}{3}a^2}}+\Phi'(a)\,\sqrt{\frac{\Lambda}{3}a^2-
\frac{m(a)}{a}} \right) \frac{\sigma}{2} \;,
\end{eqnarray}
where $\bar{p}(a)$ and $p(a)$ are the radial pressures acting on
the shell from the exterior and the interior. Equation
(\ref{pressurebalance}) relates the difference of the radial
pressure across the shell in terms of a combination of the surface
stresses, $\sigma$ and ${\cal P}$, and the geometrical quantities.
Note that $\bar{p}(a)=0$ for the exterior vacuum solution. Thus,
for the particular case of a dust shell, ${\cal P}=0$, Eq.
(\ref{pressurebalance}) reduces to
\begin{equation}
p(a)=\frac{\sigma}{2a\zeta}\;\frac{\left[(\zeta-1)+\frac{M}{a}-
(3\zeta-2)\frac{\Lambda}{3}a^2\right]}{\sqrt{1-\frac{2M}{a}-\frac{\Lambda}{3}a^2}}
\,.
\end{equation}

\section{Energy conditions on the junction}

The weak energy condition (WEC) on the junction surface implies
$\sigma \geq 0$ and $\sigma + {\cal P} \geq 0$, and by continuity
implies the null energy condition (NEC), $\sigma + {\cal P}\geq
0$. The strong energy condition (SEC) at the junction surface
reduces to $\sigma+{\cal P}\geq 0$ and $\sigma + 2{\cal P}\geq 0$,
and by continuity implies the NEC, but not necessarily the WEC.
The dominant energy condition (DEC) implies $\sigma \geq 0$ and
$\sigma \geq |{\cal P}|$.

In principle, by taking the limit $a\rightarrow r_0$ (note,
however, that $a>\bar{r}_b$), the ``total amount'' of the energy
condition violating matter, of the interior solution, may be made
arbitrarily small. Thus, in the spirit of further minimizing the
usage of exotic matter, we shall find regions where the energy
conditions are satisfied at the junction surface. For the specific
case of a dust shell, this amounts to finding regions where
$\sigma \geq 0$ is satisfied. This shall be done for the
Schwarzschild spacetime, $\Lambda=0$, the Schwarzschild-de Sitter
solution, $\Lambda >0$, and the Schwarzschild-anti de Sitter
spacetime, $\Lambda <0$. In the analysis that follows we shall
only be interested in positive values of $M$.

\subsection{Schwarzschild spacetime}

For the Schwarzschild spacetime, $\Lambda=0$, one needs to impose
that $m(r)<0$, so that Eq. (\ref{zetarestriction}) reduces to
\begin{equation}
\zeta \,
\sqrt{\frac{|m(a)|}{a}}=\frac{1-\frac{M}{a}}{\sqrt{1-\frac{2M}{a}}}
\,.
\end{equation}
The right hand term is positive, which implies that the redshift
parameter is always positive, $\zeta >0$.

Equation (\ref{sigma}) reduces to
\begin{equation}\label{Schwsigma}
\sigma=\frac{1}{4\pi a \,\zeta}
\,\left[\frac{(1-\zeta)+(\zeta-\frac{1}{2})\,\frac{2M}{a}}{\sqrt{1-\frac{2M}{a}}}
\right] \,.
\end{equation}
To analyze this relationship, we shall define a new dimensionless
parameter, $\xi=2M/a$. Therefore, Eq. (\ref{Schwsigma}) may be
rewritten in the following compact form
\begin{equation}\label{CompactSchwsigma}
\sigma=\frac{1}{8\pi M} \,\frac{\Sigma(\xi,\zeta)}{\sqrt{1-\xi}}
\,,
\end{equation}
with $\Sigma(\xi,\zeta)$ given by
\begin{equation}\label{SchwarzSigma}
\Sigma(\xi,\zeta)=\frac{1}{\zeta}\;\left[(1-\zeta)\xi+\left(\zeta-\frac{1}{2}\right)\,\xi^2
\right] \,.
\end{equation}
Equation (\ref{SchwarzSigma}) is depicted in Fig. 1. In the
interval $0< \zeta \leq 1$, we verify that $\Sigma(\xi,\zeta)>0$
for $\forall \,\xi$, implying a positive energy density, $\sigma
>0$, and thus satisfying all of the energy conditions.

For $\zeta>1$, a boundary surface, $\Sigma(\xi,\zeta)=0$, is given
at $\xi=(\zeta-1)/(\zeta-1/2)$. A positive surface energy density
is verified in the following region
\begin{equation}\label{Schwregion}
\frac{\zeta-1}{\zeta-\frac{1}{2}} \leq \xi <1 \,,
\end{equation}
i.e., $2M<a<2M(\zeta-1/2)/(\zeta-1)$.

\begin{figure}[h]
  \centering
  \includegraphics[width=2.9in]{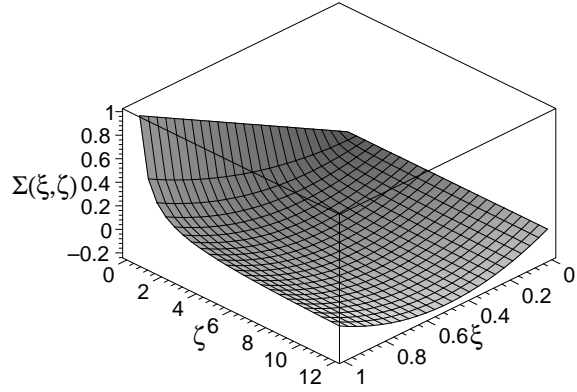}
  \caption{Plot representing the sign of the surface energy
  density. The surface is given by Eq. (\ref{SchwarzSigma}). We have
  considered the definition $\xi=2M/a$. For $0< \zeta \leq 1$ and
  $\forall \,\xi$, we verify that $\Sigma(\xi,\zeta)>0$.
  For $\zeta>1$, with $\xi=(\zeta-1)/(\zeta-1/2)$, we have
  $\Sigma(\xi,\zeta)>0$. See text for details.}
\end{figure}

\subsection{Schwarzschild-de Sitter spacetime}

For the Schwarzschild-de Sitter spacetime, $\Lambda >0$, consider
the definitions of the dimensionless parameters $\beta=9\Lambda
M^2$ and $\xi=2M/a$. Then Eq. (\ref{zetarestriction}) takes the
form
\begin{equation}
\zeta \,
\sqrt{\frac{\Lambda}{3}a^2-\frac{m(a)}{a}}=\frac{1-\frac{\xi}{2}
-\frac{8\beta}{27\xi^2}}{\sqrt{1-\xi-\frac{4\beta}{27\xi^2}}} \,.
\end{equation}
We verify that the redshift parameter is null, $\zeta=0$, if the
factor $f(\xi,\beta)=1-\frac{\xi}{2} -\frac{8\beta}{27\xi^2}$ is
zero, i.e., $\beta_n=27\xi^2(1-\xi/2)/8$, which is represented in
Fig. 2. For $\zeta >0$, we have $f(\xi,\beta)>0$, or $\beta
<\beta_n$; for $\zeta <0$, we have $f(\xi,\beta)<0$, or $\beta
>\beta_n$. Only the region below the solid
curve, given by $\beta_r=27\xi^2(1-\xi)/4$ is of interest.

The surface energy density, Eq. (\ref{sigma}), as in the previous
case, may be rewritten in the following compact form
\begin{equation}\label{SdSsigma}
\sigma=\frac{1}{8\pi M}
\,\frac{\Sigma(\xi,\zeta,\beta)}{\sqrt{1-\xi-\frac{4\beta}{27\xi^2}}}
\,,
\end{equation}
with $\Sigma(\xi,\zeta,\beta)$ given by
\begin{equation}\label{SigmaSdS}
\Sigma(\xi,\zeta,\beta)=\frac{1}{\zeta}\,\left[(1-\zeta)\xi+(\zeta-1/2)\,
\xi^2-(2-\zeta)\frac{4\beta}{27\xi}\right] \,,
\end{equation}
with $\zeta \neq 0$.

To analyze Eq. (\ref{SigmaSdS}) consider a null surface energy
density, $\sigma=0$, i.e., $\Sigma(\xi,\zeta,\beta)=0$, so that we
deduce the relationship
\begin{equation}
\beta_0=\frac{27}{4}\frac{\xi^2}{(2-\zeta)}\left[(1-\zeta)+\left(\zeta-\frac{1}{2}\right)\,\xi\right]
\,,
\end{equation}
for $\zeta \neq 2$. For the particular case of $\zeta = 2$,
depicted in Fig. 2, Eq. (\ref{SigmaSdS}) reduces to
$\Sigma(\xi,\zeta=2,\beta)=(3\xi/2-1)\xi/2$, which is null for
$\xi=2/3$, i.e., $a=3M$; positive for $\xi>2/3$, i.e., $r_b<a<3M$;
and negative for $\xi<2/3$, i.e., $3M<a<r_c$. Thus the energy
conditions are satisfied in the region to the right of the curve,
$\zeta=2$, depicted in Fig. 2.

In the interval, $0< \zeta<2$, we verify that
$\Sigma(\xi,\zeta,\beta) \geq 0$, for $\beta \leq \beta_0$. For
$\zeta>2$, then a non-negative surface energy density is verified
for $\beta \geq \beta_0$. The particular case of $\zeta=3$ is
depicted in Fig. 2. The energy conditions are also satisfied to
the right of the respective curve and to the left of the solid
line, $\beta_r$.

\begin{figure}[h]
  \centering
  \includegraphics[width=2.4in]{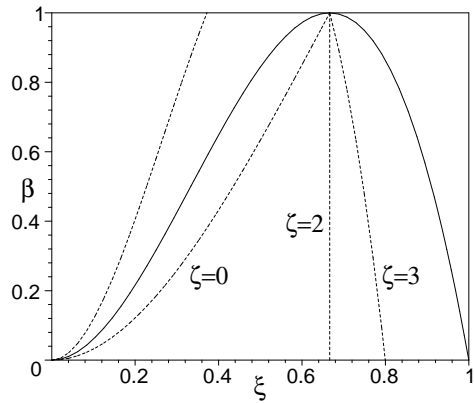}
  \caption{Analysis of the energy conditions for the
  Schwarzschild-de Sitter spacetime. We have considered the
  definitions $\beta=9 \Lambda M^2$ and $\xi=2M/a$.
  Only the region below the solid line is of interest.
  The energy conditions are
  obeyed to the right of each respective dashed curves,
  $\zeta=0$,  $\zeta=2$ and $\zeta=3$.
  See text for details.}
\end{figure}

\subsection{Schwarzschild-anti de Sitter spacetime }

For the Schwarzschild-anti de Sitter spacetime, $\Lambda <0$,
consider the parameters $\kappa=9|\Lambda| M^2$ and $\xi=2M/a$.
Then Eq. (\ref{zetarestriction}) takes the form
\begin{equation}\label{zetarestrictionSadS}
\zeta \,
\sqrt{\frac{\Lambda}{3}a^2-\frac{m(a)}{a}}=\frac{1-\frac{\xi}{2}
+\frac{4\kappa}{27\xi^2}}{\sqrt{1-\xi+\frac{4\kappa}{27\xi^2}}}
\,.
\end{equation}
The right hand side of Eq. (\ref{zetarestrictionSadS}) is always
positive, so that the restriction $\zeta>0$ is imposed.

The surface energy density, Eq. (\ref{sigma}), is given in the
following compact form
\begin{equation}\label{SadSsigma}
\sigma=\frac{1}{8\pi M}
\,\frac{\Sigma(\xi,\zeta,\kappa)}{\sqrt{1-\xi+\frac{4\kappa}{27\xi^2}}}
\,,
\end{equation}
with $\Sigma(\xi,\zeta,\kappa)$ given by
\begin{equation}\label{SigmaSadS}
\Sigma(\xi,\zeta,\kappa)=\frac{1}{\zeta}\,\left[(1-\zeta)\xi
+(\zeta-1/2)\,\xi^2+(2-\zeta)\frac{4\kappa}{27\xi} \right] \,.
\end{equation}
Consider a null surface energy density, $\sigma=0$, i.e.,
$\Sigma(\xi,\zeta,\kappa)=0$, so that from Eq. (\ref{SigmaSadS}),
we have the relationship
\begin{equation}
\kappa_0=\frac{27}{4}\frac{\xi^2}{(2-\zeta)}\left[(\zeta-1)-\left(\zeta-\frac{1}{2}\right)\,\xi\right]
\,,
\end{equation}
for $\zeta \neq 2$. For the particular case of $\zeta = 2$, Eq.
(\ref{SigmaSadS}) takes the form
$\Sigma(\xi,\zeta=2,\kappa)=(1-3\xi/2)\xi/2$, which is null for
$\xi=2/3$, i.e., $a=3M$; positive for $\xi>2/3$, i.e., $r_b<a<3M$;
and negative for $\xi<2/3$, i.e., $a>3M$. Only the region to the
left of the solid curve, given by $\kappa_r=27\xi^2(\xi-1)/4$, is
of interest.

For $0< \zeta \leq 1$, a non-negative surface energy density is
given for $\forall \, \kappa$ and $\forall \, \xi$. for $1< \zeta
<2$, $\Sigma(\xi,\zeta,\kappa)>0$ for $\kappa > \kappa_0$. For
$\zeta >2$, we have $\Sigma(\xi,\zeta,\kappa)>0$ for $\kappa_r
<\kappa < \kappa_0$.

\begin{figure}[h]
  \centering
  \includegraphics[width=2.4in]{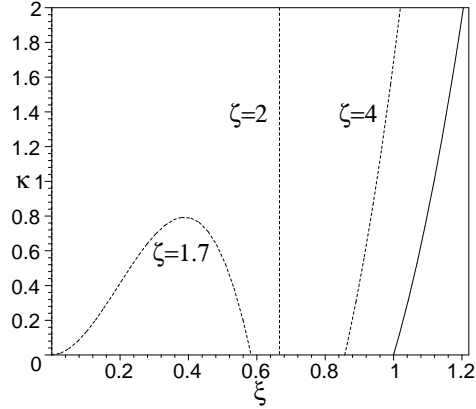}
  \caption{Analysis of the energy conditions for the
  Schwarzschild-anti de Sitter spacetime. We have considered the
  definitions $\kappa=9 |\Lambda| M^2$ and $\xi=2M/a$.
  Only the region to the left of the solid curve,
  given by $\kappa_r=27\xi^2(\xi-1)/4$, is of interest.
  For the cases of $\zeta=2$ and $\zeta=4$, the energy conditions
  are satisfied to the right of the respective dashed curves,
  and to the left of the solid line. For the specific case of
  $\zeta=1.7$, the energy conditions are obeyed above the
  respective curve. See text for details.}
\end{figure}

\section{Traversability conditions}

In this section we shall consider the traversability conditions
required for the traversal of a human being through the wormhole,
and consequently determine specific dimensions for the wormhole.
Specific cases for the traversal time and velocity will also be
estimated. In this section we shall insert $c$ to aid us in the
computations.

Consider the redshift function given by $\Phi(r)=kr^{\alpha}$,
with $\alpha, k\in \mathbb{R}$. Thus, from the definition of
$\zeta=1+a\Phi'(a)$, the redshift function, in terms of $\zeta$,
takes the following form
\begin{equation}\label{redshift}
\Phi(r)=\frac{\zeta-1}{\alpha}\,\left(\frac{r}{a}\right)^{\alpha}
,
\end{equation}
with $\alpha \neq 0$. With this choice of $\Phi(r)$, $\zeta$ may
also be defined as $\zeta=1+\alpha \Phi(a)$. The case of
$\alpha=0$ corresponds to the constant redshift function, so that
$\zeta=1$. If $\alpha<0$, then $\Phi(r)$ is finite throughout
spacetime and in the absence of an exterior solution we have
$\lim_{r\rightarrow \infty} \Phi(r)\rightarrow 0$. As we are
considering a matching of an interior solution with an exterior
solution at $a$, then it is also possible to consider the
$\alpha>0$ case, imposing that $\Phi(r)$ is finite in the interval
$r_0 \leq r \leq a$.

One of the traversability conditions required is that the
acceleration felt by the traveller should not exceed Earth's
gravity \cite{Morris}. Consider an orthonormal basis of the
traveller's proper reference frame, $({\bf e}_{\hat{0}'},{\bf
e}_{\hat{1}'},{\bf e}_{\hat{2}'},{\bf e}_{\hat{3}'})$, given in
terms of the orthonormal basis vectors of Eqs.
(\ref{basisvectors}) of the static observers, by a Lorentz
transformation, i.e.,
\begin{eqnarray}
{\bf e}_{\hat{0}'}=\gamma \,{\bf e}_{\hat{t}}\mp \gamma \,v\,{\bf
e}_{\hat{r}}  \;,
\qquad
{\bf e}_{\hat{1}'}=\mp \,\gamma \,{\bf e}_{\hat{r}} + \gamma
\,v\,{\bf e}_{\hat{t}}  \;,
\qquad
{\bf e}_{\hat{2}'}={\bf e}_{\hat{\theta}}  \;,\qquad {\bf
e}_{\hat{3}'}={\bf e}_{\hat{\phi}}  \;,
\end{eqnarray}
where $\gamma=(1-v^2)^{1/2}$, and $v(r)$ being the velocity of the
traveller as he/she passes $r$, as measured by a static observer
positioned there \cite{Morris}. Thus, the traveller's
four-acceleration expressed in his proper reference frame, ${\cal
A}^{\hat{\mu}'}=U^{\hat{\nu}'} U^{\hat{\mu}'}_{\;\;\;;
\hat{\nu}'}$, yields the following restriction
\begin{equation}\label{travellergravity}
\Bigg| \left(\frac{\Lambda}{3}r^2-\frac{m(r)}{r} \right)^{1/2}
\;e^{-\Phi}\,(\gamma e^{\Phi})'\,c^2 \Bigg| \leq g_{\oplus}  \,,
\end{equation}
The condition is immediately satisfied at the throat,
$m(r_0)=\Lambda r_0^3/3$. From Eq. (\ref{travellergravity}), one
may also find an estimate for the junction surface, $a$. Consider
that the dust shell is placed in an asymptotically flat region of
spacetime, so that $(\Lambda a^2/3-m(a)/a)^{1/2}\approx 1$. We
also assume that the traversal velocity is constant, $v={\rm
const}$, and non-relativistic, $\gamma \approx 1$. Taking into
account Eq. (\ref{redshift}), from Eq. (\ref{travellergravity})
one deduces $a \geq |\zeta-1|c^2/g_{\oplus}$. Considering the
equality case, one has
\begin{equation}\label{equalitycase2}
a = \frac{|\zeta-1|c^2}{g_{\oplus}}  \,.
\end{equation}
Providing a value for $|\zeta-1|$, one may find an estimate for
$a$. For instance, considering that $|\zeta-1|\simeq 10^{-10}$,
one finds that $a \approx 10^6 \,{\rm m}$.

Another of the traversability conditions required is that the
tidal accelerations felt by the traveller should not exceed the
Earth's gravitational acceleration \cite{Morris}. The tidal
acceleration felt by the traveller is given by $\Delta {\cal
A}^{\hat{\mu}'}=-R^{\hat{\mu}'}_{\;\;\hat{\nu}'\hat{\alpha}'\hat{\beta}'}
\,U^{\hat{\nu}'}\eta^{\hat{\alpha}'}U^{\hat{\beta}'}c^2$, where
$U^{\hat{\mu}'}=\delta^{\hat{\mu}'}_{\;\;\hat{0}'}$ is the
traveller's four velocity and $\eta^{\hat{\alpha}'}$ is the
separation between two arbitrary parts of his body. Note that
$\eta^{\hat{\alpha}'}$ is purely spatial in the traveller's
reference frame, as $U^{\hat{\alpha}'}\eta_{\hat{\alpha}'}=0$, so
that $\eta^{\hat{0}'}$. For simplicity, assume that
$|\eta^{\hat{i}'}|\approx 2\,{\rm m}$ along any spatial direction
in the traveller's reference frame \cite{Morris}. Thus, the
constraint $|\Delta {\cal A}^{\hat{\mu}'}|\leq g_{\oplus}$
provides the following inequalities
\begin{eqnarray}
&&\Bigg|\left(\frac{\Lambda}{3}r^2-\frac{m}{r}\right)
\left[\Phi''+(\Phi')^2\right] -
\frac{\Phi'}{2r^2}\left(m'r-m-\frac{2\Lambda}{3} r^3\right) \Bigg|
\, \big|\eta^{\hat{1}'}\big|\,c^2 \leq g_{\oplus}  \,,
    \label{radialtidalconstraint}
\\
&&\Bigg|\frac{\gamma^2}{2r^3}\left[\left(\frac{v}{c}\right)^2
\;\left(m'r-m-\frac{2\Lambda}{3}r^3  \right)+2r^2
\left(\frac{\Lambda}{3}r^2-\frac{m}{r}\right)\Phi' \right]
\Bigg|\, \big|\eta^{\hat{2}'}\big|\,c^2 \leq g_{\oplus}     \,.
\label{lateraltidalconstraint}
\end{eqnarray}
The radial tidal constraint, Eq. (\ref{radialtidalconstraint}),
constrains the redshift function, and the lateral tidal
constraint, Eq. (\ref{lateraltidalconstraint}), constrains the
velocity with which observers traverse the wormhole. At the
throat, $r=r_0$ or $m(r_0)=\Lambda r_0^3/3$, and taking into
account Eq. (\ref{redshift}), we verify that Eq.
(\ref{radialtidalconstraint}) reduces to
$|\eta^{\hat{1}'}\,c^2(m'-\Lambda r_0^2 )\Phi'(r_0)/2r_0| \leq
g_{\oplus}$ or
\begin{equation}\label{tidalrestriction}
\Bigg|\frac{(m'-\Lambda r_0^2)(\zeta-1)}{2r_0^2}\,
\left(\frac{r_0}{a} \right)^{\alpha} \Bigg| =
\frac{g_{\oplus}}{|\eta^{\hat{1}'}|\,c^2}  \,,
\end{equation}
considering the equality case. From this relationship, one may
find estimates for the junction interface radius, $a$, and the
wormhole throat, $r_0$. From Eq. (\ref{tidalrestriction}), one
deduces
\begin{equation}\label{equalitycase}
a = \left(\frac{|m'-\Lambda r_0^2
|\,|\zeta-1|\,|\eta^{\hat{1}'}|\,c^2}
{2g_{\oplus}r_0^2}\right)^{1/\alpha}\;r_0 \,.
\end{equation}

Using Eqs. (\ref{equalitycase2}) and (\ref{equalitycase}), one may
find an estimate for the throat radius, by providing a specific
value for $\alpha$. For instance, considering $\alpha=-1$ and
equating Eqs. (\ref{equalitycase2}) and (\ref{equalitycase}), one
finds
\begin{equation}\label{r0}
r_0=\left(\frac{|m'-\Lambda r_0^2 |\,|\zeta-1|^2 \,
|\eta^{\hat{1}'}|\,c^4}{2g_{\oplus}^2}\right)^{1/3} \,.
\end{equation}
Taking into account Eq. (\ref{flarecondition}), if $m'(r_0)\approx
\Lambda r_0^2$, we verify that the embedding diagram flares out
very slowly, so that from Eq. (\ref{r0}), $r_0$ may be made
arbitrarily small. Nevertheless, using specific examples of the
form function, for instance Eq. (\ref{Homerform}), we will assume
the approximation $|m'-\Lambda r_0^2|\approx 1$. Using the above
value of $|\zeta-1|=10^{-10}$, from Eq. (\ref{r0}) we find $r_0
\simeq 10^4\,{\rm m}$.

One may use the lateral tidal constraint, Eq.
(\ref{lateraltidalconstraint}), to find an upper limit of the
traversal velocity, $v$. Evaluated at $r_0$, we find
\begin{equation}
v \leq \sqrt{\frac{2g_{\oplus}}{|m'-\Lambda r_0^2
|\,|\eta^{\hat{2}'}|}}  \;\; r_0  \,.
\end{equation}
Taking into account the values and approximations considered
above, we have the upper bound of $v\lesssim 3\times 10^4 \,{\rm
m/s}$.

The traversal times as measured by the traveller and an observer
situated at a space station, which we shall assume rests just
outside the junction surface, are given respectively by
\cite{Morris}
\begin{eqnarray}
\Delta \tau=\int_{-a}^a\,\frac{dl}{v \gamma}
\qquad \hbox{and} \qquad
\Delta t=\int_{-a}^a\,\frac{dl}{v e^{\Phi}} \,,
\end{eqnarray}
where $dl=\left(\Lambda r^2/3-m/r \right)^{-1/2}dr$ is the proper
radial distance. Since we have chosen $\gamma \approx 1$ and
$|\zeta -1|\approx 10^{-10}$, we can use the following
approximations
\begin{equation}
\Delta \tau \approx \Delta t \approx \int_{-a}^a\,\frac{dl}{v}
\approx \frac{2a}{v}  \,.
\end{equation}
For instance, considering the maximum velocity, $v\approx 3\times
10^4 \,{\rm m/s}$, with $a\approx 10^6\,{\rm m}$, the traversal
through the wormhole can be made in approximately a minute.

\section{Conclusion}
We have presented a specific metric of a spherically symmetric
traversable wormhole in the presence of a generic cosmological
constant, verifying that the null energy condition and the
averaged null energy condition are violated, as was to be
expected. We verified that evaluating the ``volume integral
quantifier'', the specific Ellis drainhole may also be
theoretically constructed with arbitrarily small quantities of
averaged null energy condition violating matter. Furthermore, we
constructed a pressureless dust shell around the interior wormhole
solution, by matching the latter to a unique vacuum exterior
spacetime, in the presence of a generic cosmological constant. In
the spirit of minimizing the usage of exotic matter, regions were
determined in which the surface energy density is non-negative,
thus satisfying all of the energy conditions at the junction
surface. An equation governing the behavior of the radial pressure
across the shell was also determined. Taking advantage of the
construction, and considering the traversability conditions,
estimates of the throat radius and the junction interface radius,
of the total traversal time and maximum velocity of an observer
journeying through the wormhole, were also found.

\end{document}